\documentclass[aps,prc,twocolumn,showpacs,floatfix,nofootinbib,superscriptaddress,amsmath,amssymb,longbibliography,amsproc]{revtex4-2}

\usepackage{animate}
\usepackage{graphicx}
\usepackage{amsmath}
\usepackage{amssymb}%
\usepackage{dcolumn}
\usepackage{multirow}
\usepackage{listings}
\usepackage[dvipsnames]{xcolor}
\usepackage[colorlinks=true,linkcolor=blue,citecolor=blue,urlcolor=blue]{hyperref}
\usepackage{todonotes}
\usepackage{orcidlink}

\graphicspath{{.}{./figures/}}
\begin{document}

\title{Dynamics of density fluctuations in atomic nuclei}

\author{Francesca~Bonaiti\orcidlink{0000-0002-3926-1609}}
\email{bonaiti@frib.msu.edu}
\affiliation{Facility for Rare Isotope Beams, Michigan State University, East Lansing, Michigan 48824, USA}
\affiliation{Physics Division, Oak Ridge National Laboratory, Oak Ridge, Tennessee 37831, USA}

\author{Gaute~Hagen\orcidlink{0000-0001-6019-1687}}
\affiliation{Physics Division, Oak Ridge National Laboratory, Oak Ridge, Tennessee 37831, USA}
\affiliation{Department of Physics and Astronomy, University of Tennessee, Knoxville, Tennessee 37996, USA}

\author{Thomas~Papenbrock\orcidlink{0000-0001-8733-2849}}
\affiliation{Department of Physics and Astronomy, University of Tennessee, Knoxville, Tennessee 37996, USA}
\affiliation{Physics Division, Oak Ridge National Laboratory, Oak Ridge, Tennessee 37831, USA}

\begin{abstract}
We study the spatiotemporal patterns of density fluctuations in  $^{16,24}$O and $^{48}$Ca using nuclear interactions from chiral effective field theory and the time-dependent coupled-cluster method. We find that two-particle--two-hole excitations generate small-amplitude fluctuations that are fast, short-ranged and of stochastic character.   
\end{abstract}
\maketitle

\textit{Introduction.---} Nuclear dynamics involves fascinating processes such as fission where a heavy nucleus splits into lighter fragments. The time evolution of this process is complicated because it involves vastly different time scales and complex quantum many-body states, see Refs.~\cite{simenel2018,sekizawa2019,schunck2022} for recent reviews. The inverse process, fusion, is similarly complicated when heavy nuclei are involved and plays a key role in the synthesis of superheavy elements. 

The dynamics of fission and fusion is usually modeled with time-dependent Hartree Fock (Bogoliubov) methods and density functional theory. The recent computations~\cite{goddard2015,bulgac2016,sekizawa2016,goddard2020,hasegawa2020,abdurrahman2024} revealed that the time scales involved in fission typically range from tens to hundreds of fm$/c$, with equilibration being slowest and of the order of $1~\mathrm{zs}=10^{-21}~\mathrm{s}\approx 300~\mathrm{fm}/c$. Similarly, relevant experimental timescales for nuclear reactions are also of the order of 0.3~zs~\cite{jedele2017,simenel2020}.

These timescales are clearly visible in movies about nuclear collisions, as shown, e.g., by \textcite{sekizawa2016}. Their movies show how densities evolve in time. These were produced at about 30~frames per second, and a second of showtime corresponds to $60~\mathrm{fm}/c$. Thus, the time step between frames is about 2~fm$/c$, a numerical sampling interval much shorter than the time scales over which the densities change. One sees colliding nuclei that  -- dependent on the impact parameter -- fuse or separate again, with the reaction products slowly vibrating and/or rotating.  

How does nuclear dynamics look in an \textit{ab initio} approach~\cite{ekstrom2023}? Hamiltonians from effective field theories of quantum chromodynamics typically have momentum cutoffs 
of 400--500~MeV$/c$. This corresponds to nucleon kinetic energies that are about 100~MeV, and relevant physical time scales that could be as short as 2~fm$/c$, much shorter than what is observed in time-dependent mean-field computations. Furthermore, \textit{ab initio} computations include two-particle--two-hole excitations and thereby differ from the mean-field methods that are limited to one-particle--one-hole excitations. So, one clearly expects shorter time scales and more complex dynamics in an \textit{ab initio} approach.  

In this Letter, we study basic aspects of nuclear dynamics in an \textit{ab initio} setting. The long-term goal is to compute fission and fusion within this framework but that requires further advances. Time-independent \textit{ab initio} computations have reached the spherical nuclei $^{208,266}$Pb~\cite{hu2022,bonaiti2025a} and deformed nuclei in the mass $A\approx 80$ region~\cite{hu2024}. In contrast, time-dependent \textit{ab initio} computations of nuclei, however, are less advanced~\cite{pigg2012}. One reason was the insufficient computational resources available in the past. In time-dependent coupled-cluster theory~\cite{hoodbhoy78,hoodbhoy79,kvaal2012,sverdrupofstad2020}, for example, one has to solve coupled-cluster equations for each time step; as we will see below, time steps are of the order of 0.2~fm$/c$, and studying a single time evolution to hundreds of  
fm$/c$ thus becomes a thousand times more expensive than a single static solution. However, present-day computing resources are now making this possible~\cite{bonaiti2025b}. 

We want to study the dynamics of nuclear density fluctuations. We are interested in their time scales and their amplitudes. 
To do so, we use  time-dependent coupled-cluster theory as formulated in Refs.~\cite{kvaal2012,pigg2012}. 
Our approach is based on Hamiltonians from chiral effective field theory~\cite{epelbaum2009,machleidt2011,hebeler2011,hammer2020}.  Our time evolutions start from densities that resulted from  time-independent Hartree-Fock computations of the nucleus under consideration. Such densities are close to those obtained from more accurate coupled-cluster computations. (Their charge radii, for instance, differ only on the percent level~\cite{koszorus2021,companys2025}.) We then let such a density evolve in time, solving the time-dependent Schr\"odinger equation with the coupled-cluster method. As we will see, the density oscillates around the density that is a solution of the time-independent coupled-cluster equations. (This density is an equilibrium fixed point in the time evolution.) On time scales of tens of fm$/c$, the oscillations are qualitatively similar to what is seen in the movies of Ref.~\cite{sekizawa2016}. However, in addition to these oscillations there are also short-time fluctuations, which are small in amplitude, of short range, and somewhat stochastic. This is the exciting discovery we report in this Letter.

\textit{Time-dependent coupled cluster theory.---}
Our coupled-cluster computations start from a Hartree-Fock product state 
\begin{equation}
\label{vac}
    |\phi_0\rangle \equiv \prod_{i=1}^A \hat{a}^\dagger_i|0\rangle \ .
\end{equation}
which serves as the reference state.
Here, $|0\rangle$ is the zero-particle vacuum state, and $\hat{a}_p^\dagger$ creates a nucleon in the single-particle state $|p\rangle\equiv\hat{a}_p^\dagger|0\rangle$. In Eq.~(\ref{vac}) and in what follows, we used the convention that labels $i,j,k,\ldots$ ($a,b,c,\ldots$) refer to occupied (unoccupied) states, while $p,q,r,\ldots$ label any state. 
In this work, the creation and annihilation operators are taken to be time-independent (see Refs.~\cite{kvaal2012,pigg2012,sverdrupofstad2020} for formulations with time-dependent orbitals).  

In coupled-cluster theory, the ket ground-state is written as 
\begin{equation}
\label{ket}
    |\psi(t)\rangle = e^{\hat{T}(t)}|\phi_0\rangle \ ,  
\end{equation}
where $\hat{T}(t)= \hat{T}_1(t) +\hat{T}_2(t) + \ldots +\hat{T}_A(t)$ with
\begin{align}
    \hat{T}_1(t) & \equiv \sum_{ia} t_i^a(t) \hat{a}^\dagger_a\hat{a}_i \ , \\
    \hat{T}_2(t) & \equiv \frac{1}{4}\sum_{ijab} t_{ij}^{ab}(t) \hat{a}^\dagger_a\hat{a}^\dagger_b\hat{a}_j\hat{a}_i \ , 
\end{align}
and so on. The bra ground-state is parameterized as 
\begin{equation}
\label{bra}
    \langle \widetilde{\psi}(t)| = \langle\phi_0|\left[1+\hat{L}(t)\right] e^{-\hat{T}(t)} \ .
\end{equation}
Here $\hat{L}(t)= \hat{L}_1(t) +\hat{L}_2(t) + \ldots +\hat{L}_A(t)$  is a de-excitation operator with 
\begin{align}
    \hat{L}_1(t) & \equiv \sum_{ia} l^i_a(t) \hat{a}^\dagger_i\hat{a}_a \ , \\
    \hat{L}_2(t) & \equiv \frac{1}{4}\sum_{ijab} l^{ij}_{ab}(t) \hat{a}^\dagger_i\hat{a}^\dagger_j\hat{a}_b\hat{a}_a \ ,
\end{align}
and so on. The bra ground state is not the adjoint of the ket, and the coupled-cluster equations stem from a bi-variational principle~\cite{kvaal2012,kvaal2025}. In what follows, we will consider two approximations, namely coupled-cluster singles and doubles (CCSD), where $\hat{T}=\hat{T}_1+\hat{T}_2$ and $\hat{L}=\hat{L}_1+\hat{L}_2$ are truncated at the two-particle--two-hole level, and coupled-cluster doubles (CCD), where only $\hat{T}_2$ and $\hat{L}_2$ are retained.

As we focus on density fluctuations, our interest is in the time-dependent density
\begin{equation}
    \rho(r,t) = \langle\phi_0|\left(1+\hat{L}(t)\right)e^{-\hat{T}(t)} \hat{\rho}(r) e^{\hat{T}(t)} |\phi_0\rangle \ ,
\end{equation}
where $\hat{\rho}(r)$ is the one-body density operator. 

The dynamical coupled-cluster equations can be derived from a variation of an action functional~\cite{,kvaal2012,kvaal2025} with respect to the parameters of $\hat{L}$ and $\hat{T}$. In the CCSD approximation one finds
\begin{equation}
\begin{split}
    i\hbar \dot{t}_0(t) &= \langle\Phi_0|\overline{H}|\Phi_0\rangle \ , \\
    i\hbar \dot{t}_i^a(t) &= \langle\Phi^a_i|\overline{H}|\Phi_0\rangle \ , \\
    i\hbar \dot{t}_{ij}^{ab}(t) &= \langle\Phi^{ab}_{ij}|\overline{H}|\Phi_0\rangle \ ,  
\end{split}
\label{ccsdeq}
\end{equation}  
and
\begin{equation}
\begin{split}
    -i\hbar \dot{l}_0(t) &= 0 \ , \\
    -i\hbar \dot{l}_i^a(t) &= \langle\Phi_0|L(t)\overline{H}|\Phi^a_i\rangle \ , \\
    -i\hbar \dot{l}_{ij}^{ab}(t) &= \langle\Phi_0|L(t)\overline{H}|\Phi^{ab}_{ij}\rangle \ . 
    \label{lambdaeq}
\end{split}
\end{equation}
Here, 
\begin{equation}
    \overline{H} \equiv e^{-\hat{T}(t)} H e^{\hat{T}(t)}
\end{equation}
is the similarity-transformed Hamiltonian. In the CCD approximation one sets $t_i^a=0=l^i_a$ at all times. 

Our initial condition is the Hartree-Fock state~(\ref{vac}). Thus, $t_i^a(0)=0$, $t_{ij}^{ab}(0)=0$,  $l^i_a$=0, and $l^{ij}_{ab}(0)=0$. The time evolution will then introduce changes in the density via particle-hole-excitations. 
We define density fluctuations as 
\begin{equation}
\label{key}
    \delta\rho(r,t)\equiv \rho(r,t) - \rho_{\rm av}(r) \ .
\end{equation}
Here the time-averaged density $\rho_{\rm av}(r)$ is 
\begin{equation}
\label{avdens}
        \rho_{\mathrm av}(r) \equiv \frac{1}{\tau_{\rm max} - \tau_{\rm min}}\int\limits_{\tau_{\rm min}}^{\tau_{\max}} dt\; \rho(r,t) \ .
\end{equation} 
This average is approximately independent of the integration limits $\tau_{\rm min}$ and $\tau_{\rm max}$ for sufficiently large times $\tau_{\rm min}, \tau_{\rm max}$, of the order of $100$~fm/$c$. 

We note that an equilibrium (or fixed) point of the dynamics appears when the right-hand side of the Eqs.~(\ref{ccsdeq}) and (\ref{lambdaeq}) vanishes. At this fixed point, the $\hat{T}$ and $\hat{L}$ amplitudes fulfill the time-independent coupled-cluster equations~\cite{shavittbartlett2009}. One can linearize the equations in the vicinity of the fixed point and then obtains harmonic oscillations (for which the long-time average is the fixed point). While the Hartree-Fock density is close to the fixed-point density (see Supplemental Material), one is generally not in the linear regime. Consequently, the time averaged density~(\ref{avdens}) differs slightly from the CCSD density. 

We use {\sc cvode}, an adaptive solver from the {\sc{Sundials}}~\cite{hindmarsh2005sundials,gardner2022sundials} library, to integrate the set of ordinary differential equations~(\ref{ccsdeq}--\ref{lambdaeq}). During the time evolution, we employ time steps of 0.2~fm$/c$ in {\sc cvode} and record results at increments of 1~fm$/c$. 

\textit{Hamiltonian and model space.---} In this work, we employ two chiral effective field theory interactions, NNLO$_{\rm sat}$~\cite{ekstrom2015a} and $\Delta$NNLO$_{\rm GO}$(394)~\cite{jiang2020}. These interactions are formulated at next-to-next-to-leading order in the chiral expansion, and include three-body forces. 
NNLO$_{\rm sat}$ and $\Delta$NNLO$_{\rm GO}$(394) exhibit a momentum cutoff of $450$ MeV/$c$ and $394$ MeV/$c$, respectively. Since we focus on short-time and short-range phenomena in this work, the difference in cutoff scales allows us to illustrate their model dependence. 

Our time-dependent coupled-cluster calculations are performed in a Hartree–Fock basis. The model space consists of a single-particle basis from the spherical harmonic oscillator with excitations up to and including $N_{\rm max}\hbar\omega$. We use $N_{\rm max} = 8$ and the harmonic oscillator frequency $\hbar\Omega = 16$~MeV. $N_{\rm max} = 8$ is currently the largest model space available for such computations. While the latter is not large enough to fully converge the ground-state energy, it is sufficient to reproduce collective excitations~\cite{sun2024,papenbrock2024}. We also checked that qualitatively similar results are obtained also for $N_{\rm max} = 6$ (see Supplemental Material).

\textit{Results.---}   
Figure~\ref{fig:48Ca} shows the spatial and temporal evolution of density fluctuations~(\ref{key}) for the nucleus $^{48}$Ca, computed with both interactions and using the CCSD and CCD approximations. The colors red and blue denote positive and negative density fluctuations, respectively, with white being zero. [We used $\tau_{\rm min} = 100$~fm$/c$ and $\tau_{\rm max} = 200$~fm$/c$ to compute the time-averaged density~(\ref{avdens}).] 

\begin{figure}[htp]
    \centering
    \includegraphics[width=0.49\textwidth]{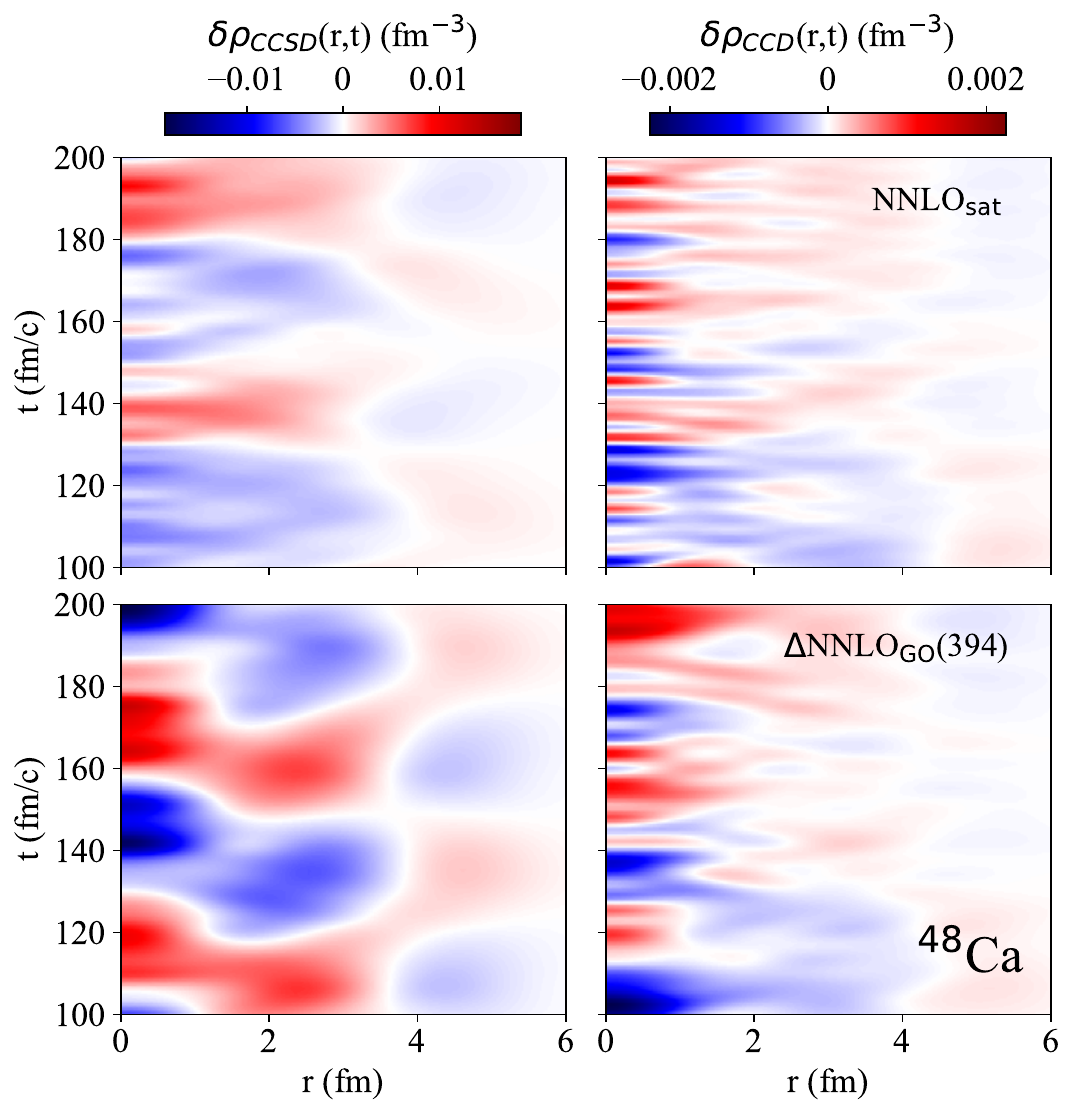}
    \caption{Upper panels: Matter density fluctuations of $^{48}$Ca with respect to the time-averaged density as a function of radial distance $r$ and time $t$ computed with the NNLO$_{\rm sat}$ interaction in the CCSD (left) and CCD (right) approximation scheme. A model space with $N_{\rm max}=8$ and $\hbar\Omega = 16$ MeV has been employed. Lower panels: same for the $\Delta$NNLO$_{\rm GO}$(394) interaction.}
    \label{fig:48Ca}
\end{figure}

The CCSD results (left column in Fig.~\ref{fig:48Ca}) exhibit slow oscillations with a period of about 50~fm$/c$ and their amplitude is about 6\% of the saturation density $\rho_0\approx0.16$~fm$^{-3}$ for small radii. These slow oscillations are present for both interactions (though they are out of phase). The amplitude of density fluctuations decrease rapidly around the nuclear surface at about 3.5~fm. The typical time scale of the slow oscillatory density fluctuations, about tens of fm$/c$, is similar to what has been seen in time-dependent Hartree-Fock simulations of fission~\cite{goddard2015,sekizawa2016,jedele2017,simenel2020,goddard2020,hasegawa2020,abdurrahman2024}. However, we also observe fringes of much shorter time scales at small radii. These are more pronounced for the NNLO$_{\rm sat}$ interaction which has a higher momentum cutoff (see top row in Fig.~\ref{fig:48Ca}). 

To better examine these fast fluctuations we also performed time-dependent CCD computations. 
The right panels of Fig.~\ref{fig:48Ca} shows the density fluctuations~(\ref{key}) from two-particle--two-hole excitations, computed with the CCD approximation. The fastest fluctuations are short-ranged, extending up to $1-2$ fm, and their amplitude is one order of magnitude smaller than the slow oscillations visible in the left panels. The short-range fluctuations happen on short time scales of about $3-4$~fm$/c$, for the NNLO$_{\rm sat}$ interaction, while they are somewhat longer for the softer $\Delta$NNLO$_{\rm GO}$(394) interaction. To put the fast time scales into perspective with nuclear reaction dynamics we note that the former are one order of magnitude smaller than those of neck rupture in mean-field computation of fission (which are of the order of $10^{-22}$~s~\cite{abdurrahman2024}) and those of equilibration in nuclear reactions~\cite{jedele2017} and collisions~\cite{simenel2020}, which are of the order of 1~zs~$= 10^{-21}$s. (We recall that 1~fm/$c\approx 3.3\times 10^{-24}$~s.)

Let us compare the shortest oscillation period observed in Fig.~\ref{fig:48Ca} with the maximum energy scale that can be resolved by the nuclear Hamiltonian, set by the momentum cutoff $\Lambda$ of the interaction. There is no advantage in increasing $\Lambda$ beyond the breakdown scale of chiral effective field theory. For the NNLO$_{\rm sat}$ interaction, the shortest period observed in Fig.~\ref{fig:48Ca} is $3$~fm/$c$, corresponding to an energy scale $E \sim \pi \hbar c /(3\,\mathrm{fm}) \approx 207$~MeV. Considering that $\Lambda = 450$ MeV/c for NNLO$_{\rm sat}$, $E$ is consistent with the maximum kinetic energy $E_{\rm kin, max}$ $= \Lambda^2/(2\mu) \approx 215$~MeV of a two-nucleon system (with reduced mass $\mu$). A softer interaction, as $\Delta$NNLO$_{\rm GO}$(394), with a smaller cutoff, yields a lower $E_{\rm kin, max}$ and consequently longer minimum oscillation periods, as can be seen in Fig.~\ref{fig:48Ca}. Therefore, the time step used in the time evolution, which in our calculations is $0.2$~fm/$c$, is chosen to be much smaller than the time scale associated with the interaction cutoff. Videos showing short-range fluctuations emerging during time evolution in $^{48}$Ca for both interactions can be found in the Supplemental Material. 

\begin{figure}[htp]
    \centering
    \includegraphics[width=0.49\textwidth]{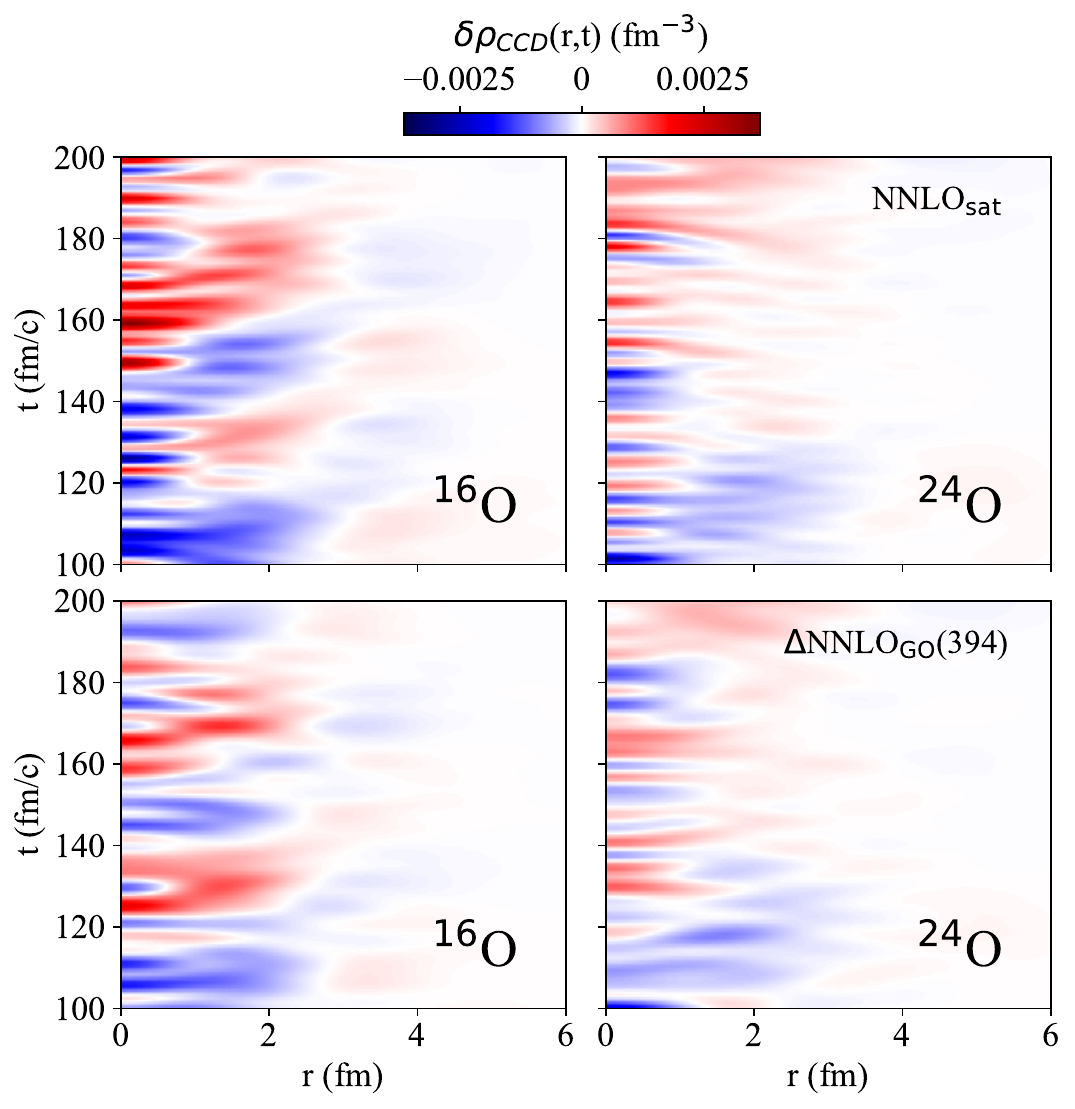}
    \caption{Upper panels: Matter density fluctuations of $^{16}$O (left) and $^{24}$O (right) with respect to the time-averaged density as a function of radial distance $r$ and time $t$ computed with the NNLO$_{\rm sat}$ interaction in the CCD approximation scheme. A model space with $N_{\rm max}=8$ and $\hbar\Omega = 16$ MeV has been employed. Lower panels: same for the $\Delta$NNLO$_{\rm GO}$(394) interaction.}
    \label{fig:oxygen}
\end{figure}

Let us analyze density fluctuations in different nuclei. In Fig.~\ref{fig:oxygen} we show CCD results for $^{16,24}$O for both the NNLO$_{\rm sat}$ and $\Delta$NNLO$_{\rm GO}$(394) interactions. In analogy to $^{48}$Ca, for both oxygen isotopes we observe faster time scales, around 3 fm/$c$, for the NNLO$_{\rm sat}$ interaction, while the $\Delta$NNLO$_{\rm GO}$(394) results feature longer periods up to $10$ fm/$c$. Short-range fluctuations exhibit a slightly larger spatial extent in $^{24}$O, consistent with its larger matter radius. It is worth pointing out that the amplitude of density fluctuations is similar in both $^{16,24}$O and $^{48}$Ca. Also in the case of $^{16,24}$O, they are one order of magnitude smaller than the corresponding CCSD result. It appears that short-time, short-range fluctuations are model dependent yet universal: The model dependence comes from the scheme and regulator dependence of short-range physics while -- for a given Hamiltonian -- the fluctuations are universal as they do not depend on the mass number.

This interpretation is consistent with recent studies of static short-range correlations in experiment~\cite{subedi2008,fomin2012,hen2014,duer2019,schmidt2020} and theory~\cite{schiavilla2007,wiringa2014,weiss2015,weiss2018,weiss2019,lynn2020,pastore2020,weiss2021,tropiano2021,tropiano2024}. In contrast to those works, however, this Letter reveals their spatial and temporal patterns simultaneously. The dominance of nucleon pairs in static short-range correlations~\cite{korover2023,beck2023} is clearly consistent with the importance of two-particle--two-hole excitations in the dynamical fluctuations. In relation to studies of static short-range correlations, it would be interesting to explore the effect of three-particle--three-hole excitations on density fluctuations, as well as their isospin dependence. We leave this for future work.

\textit{Discussion.---} We discuss our results in the context of $(i)$ beyond-mean-field simulations of nuclear dynamics and $(ii)$ stochastic behavior in nuclei. 
First, the short-range, short-time oscillations reported in this Letter are induced by two-particle two-hole correlations. As such they cannot be detected in a mean-field approach, which does not yield any fluctuations in observables~\cite{balian1984,simenel2025}. The inclusion of these quantum fluctuation requires beyond-mean-field methods~\cite{hoodbhoy78,hoodbhoy79,balian1992,simenel2025}. Examples are the beyond-mean-field approaches~\cite{goutte2005,regnier2016} to fission product distributions where collective wave packets were evolved in time, or time-dependent random phase approximation calculations~\cite{simenel2020}. One can also introduce stochastic aspects~\cite{abe1996,frobich1998} by employing an ensemble of initial conditions~\cite{ayik2008}.
In contrast, time-dependent coupled-cluster theory provides us with a systematic expansion to include many-body correlations beyond the mean field.  

\begin{figure}[htp]
    \centering
    \includegraphics[width=0.49\textwidth]{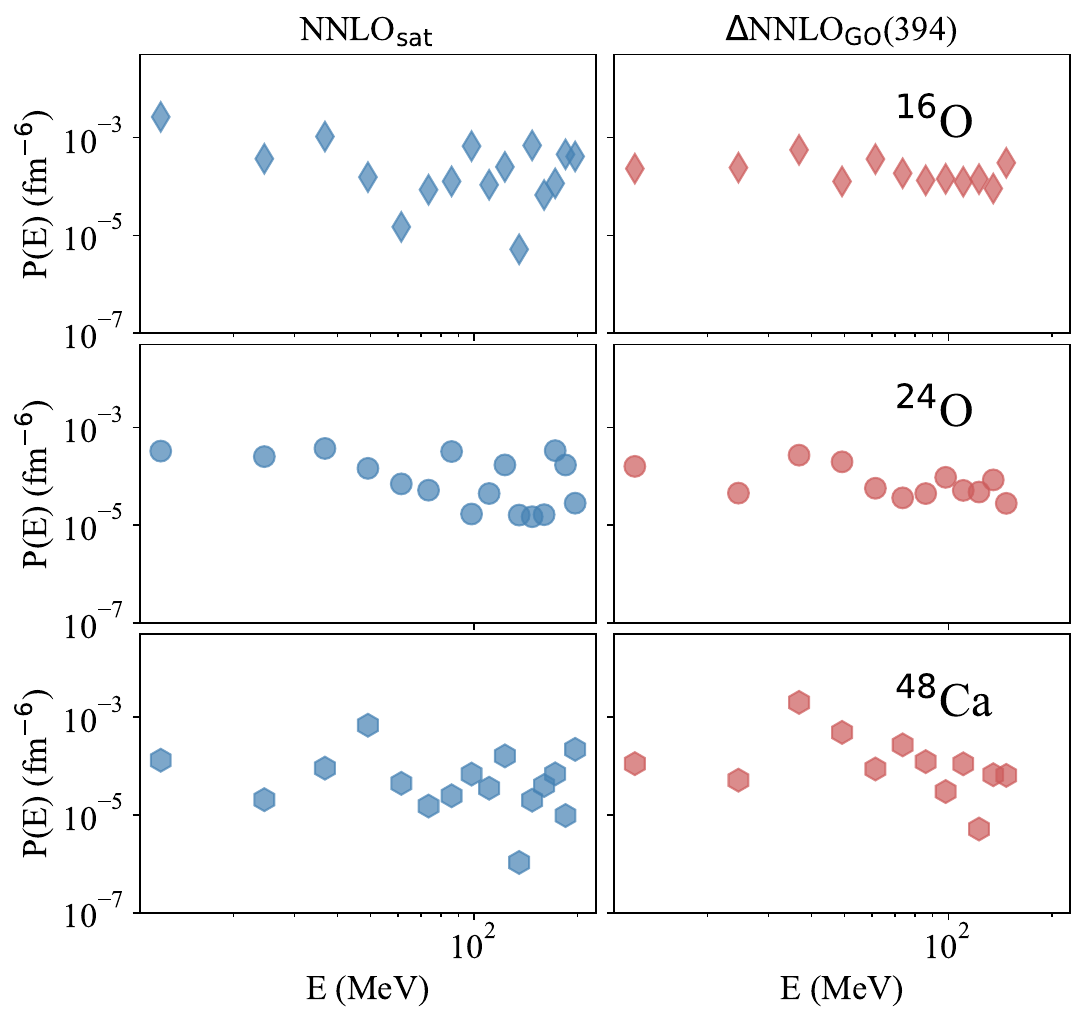}
    \caption{Left panels: Power spectrum of CCD density fluctuations in $^{16}$O (diamonds), $^{24}$O (circles) and $^{48}$Ca (hexagons) computed with the NNLO$_{\rm sat}$ interaction. Right panels: same for the $\Delta$NNLO$_{\rm GO}$(394) interaction. Both axes employ a logarithmic scale.}
    \label{fig:power_spectra}
\end{figure}

Second, the observed short-range, short-time fluctuations look somewhat stochastic. To analyze this further we compute the power spectrum $P(E) = |\widetilde{\delta\rho}(E)|^2$ of the  time-dependent signal~(\ref{key}) at a fixed radial distance $r_0$ via a Fourier transform. Since our focus is on short-range effects, we choose $r_0 \approx 10^{-4}$ fm (we verified that results are qualitatively similar for other $r_0$ values below $1$~fm). Figure~\ref{fig:power_spectra} shows the results from CCD computations of $P(E)$ for $^{16,24}$O and $^{48}$Ca obtained with both the NNLO$_{\rm sat}$ and the $\Delta$NNLO$_{\rm GO}$(394) interactions. 

Here we excluded the data points that are above the kinetic energies $\Lambda^2/(2\mu)$ for momentum cutoffs $\Lambda$. (These are around $165$~MeV for $\Delta$NNLO$_{\rm GO}$(394) and $215$~MeV for NNLO$_{\rm sat}$.) 

For all nuclei, the power spectrum appears to be almost energy-independent, comparable to white noise. This suggests that the dynamics of short-range density fluctuations is intrinsically stochastic. NNLO$_{\rm sat}$ results appear to be more scattered at higher energy than those based on $\Delta$NNLO$_{\rm GO}$(394), consistent with the smaller oscillation periods observed for NNLO$_{\rm sat}$ and its larger momentum cutoff. The power spectrum is spanning approximately the same orders of magnitude for all nuclei, reflecting the universal behavior of density fluctuations. This raises the question if one could interpret the stochastic behavior of density fluctuations as a signature of chaos in atomic nuclei. Traditionally, this relation rests on the fact that random matrix theory describes the statistics of nuclear spectra and resonance widths~\cite{verbaarschot1985,zelevinsky1996,papenbrock2007,weidenmueller2009,mitchell2010,kawano2016}. The thought is that the nuclear interaction is complicated and strongly mixes shell-model states, based on the compound nucleus picture by \textcite{bohr1936} and the ideas about random matrices by \textcite{wigner1956}. Of course, random matrix theory also applies to the spectral statistics of classically chaotic systems~\cite{bohigas1984,mueller2004}. While the computation of out-of-time correlators~\cite{xu2024} might be useful to explore this connection, it is eye-opening to see stochastic aspects directly in nuclear dynamics.

\emph{Summary.--} \textit{Ab initio} computations of nuclear dynamics reveal the presence of short-range, short-time density fluctuations, which are stochastic in character. These fluctuations emerge across different nuclei and for different interactions. They are as short as $3$~fm/$c$ in typical interactions from chiral effective field theory. Their presence in nuclear dynamics is universal and unavoidable when correlations beyond the mean field are taken into account. 

\emph{Data availability.--} The data that support the findings of this article are openly available~\cite{bonaiti_density_zenodo2026}.

\acknowledgments
We thank Kyle Godbey and Chloë Hebborn for helpful discussions. This work was supported by the U.S. Department of Energy, Office of Science, Office of Nuclear Physics, under the FRIB Theory Alliance award DE-SC0013617 and Award No.~DE-FG02-96ER40963, by SciDAC-5 (NUCLEI collaboration). Computer time was provided by the Innovative and Novel Computational Impact on Theory and Experiment (INCITE) programme and by the Institute for Cyber-Enabled Research at Michigan State University. This research used resources of the Oak Ridge Leadership Computing Facility at the Oak Ridge National Laboratory, which is supported by the Advanced Scientific Computing Research programs in the Office of Science of the U.S. Department of Energy under Contract No. DE-AC05-00OR22725.

\bibliography{master}
\clearpage
\onecolumngrid
\begin{center}
\textbf{Dynamics of density fluctuations in atomic nuclei: \\Supplemental Material}
\end{center}
\twocolumngrid

In this Supplemental Material, we compare density fluctuations computed with respect to the time-averaged density and the Hartree-Fock density and we study the dependence of density fluctuations on the model space size. A short description of the videos of density fluctuations included in the Supplemental Material is also provided.  

\begin{figure}[htb]
    \centering
    \includegraphics[width=0.49\textwidth]{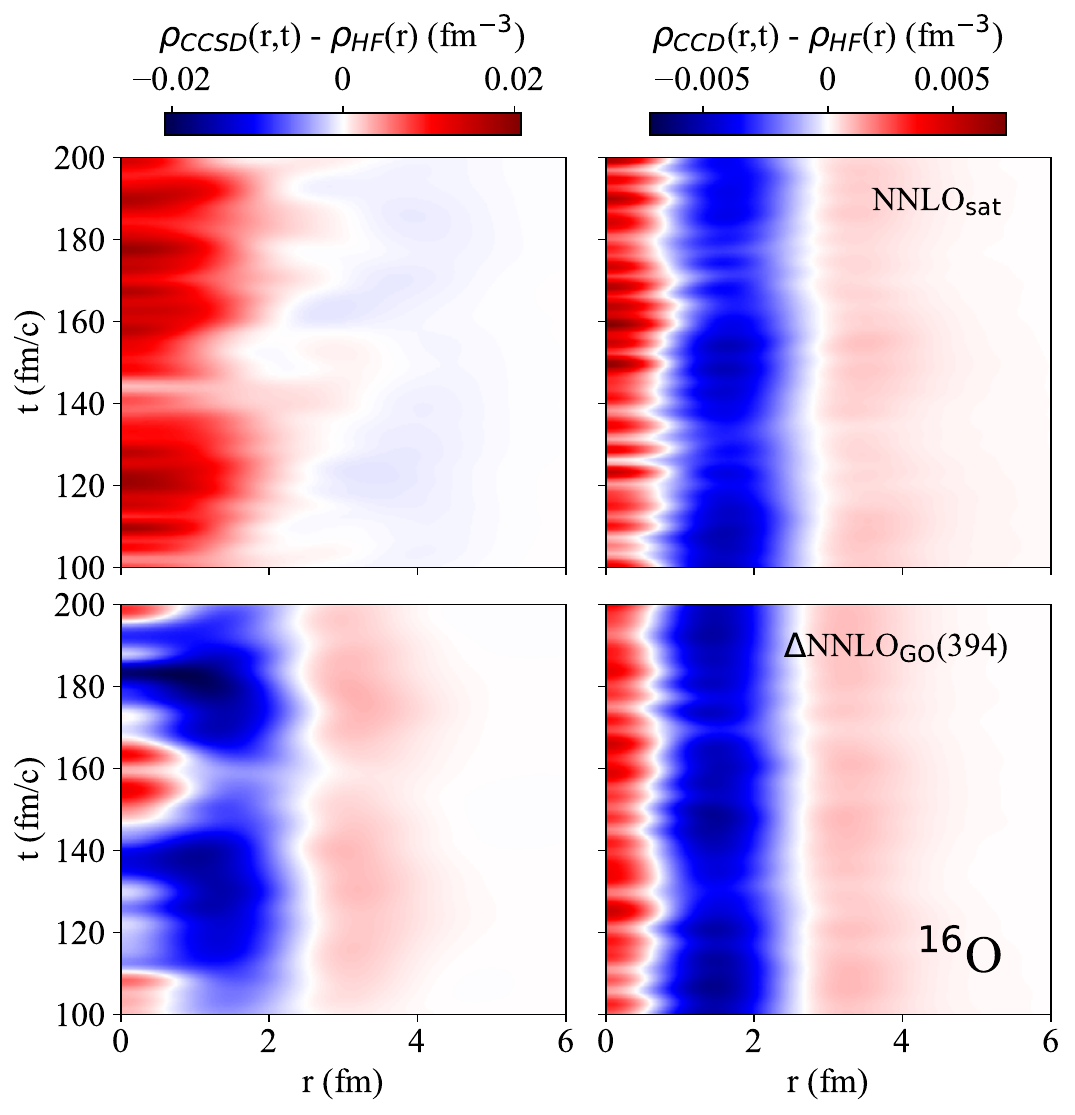}
    \caption{Upper panels: Matter density fluctuations of $^{16}$O with respect to the Hartree-Fock density as a function of radial distance $r$ and time $t$ computed with the NNLO$_{\rm sat}$ interaction in the CCSD (left) and CCD (right) approximation scheme. A model space with $N_{\rm max}=8$ and $\hbar\Omega = 16$ MeV has been employed. Lower panels: same for the $\Delta$NNLO$_{\rm GO}$(394) interaction.}
    \label{fig:16O_HF}
\end{figure}

\emph{Density fluctuations with respect to the Hartree-Fock density.-- } For the purposes of this paragraph, we focus on $^{16}$O. One could think of computing density fluctuations with respect to the initial Hartree-Fock density $\rho_{\rm HF}(r)$. In Figure~\ref{fig:16O_HF}, we show  density fluctuations with respect to the Hartree-Fock density computed within the the CCSD and CCD approximations and using the NNLO$_{\rm sat}$ and $\Delta$NNLO$_{\rm GO}$(394) interactions.

Signatures of short-time, short-range, and small amplitude fluctuations are visible in Figure~\ref{fig:16O_HF}, but these are obscured by the large (and almost static) deviation between the dynamical density and the Hartree-Fock density. The latter must clearly be quite different from the time-averaged density.  

\begin{figure}[htb]
    \centering
    \includegraphics[width=0.49\textwidth]{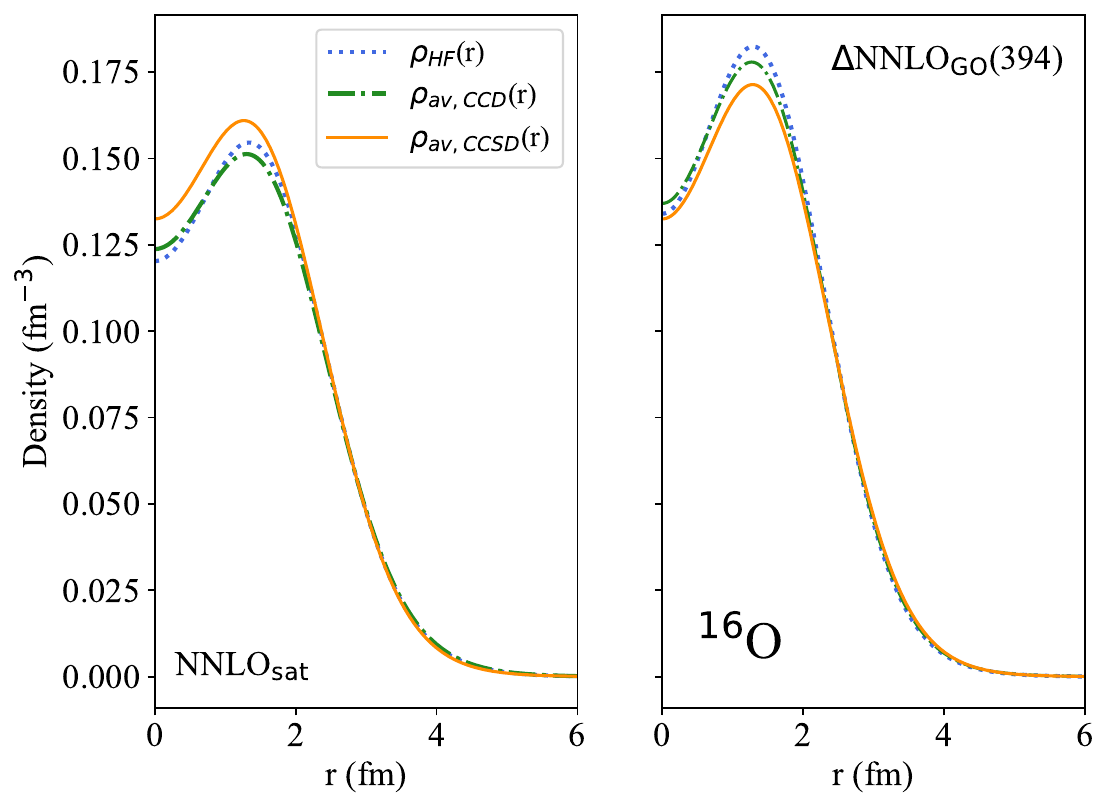}
    \caption{Comparison between the Hartree-Fock density and the time-averaged CCSD and CCD densities for the NNLO$_{\rm sat}$ (left panel) and $\Delta$NNLO$_{\rm GO}$(394) (right panel) interactions in $^{16}$O. A model space with $N_{\rm max} = 8$, $\hbar\Omega = 16$ MeV has been employed. }
    \label{fig:densities}
\end{figure}

To see this,  it is instructive to compare the Hartree-Fock density to the time-averaged densities from CCSD and CCD for both interactions. The comparison is shown in Figure~\ref{fig:densities}. 

Let us first focus on the CCSD time-averaged density $\rho_{\rm av, CCSD}(r)$. For the NNLO$_{\rm sat}$ interaction, this density is higher than $\rho_{\rm HF}(r)$ at short distances, leading to the red pattern in Figure~\ref{fig:16O_HF}. The opposite situation is observed for $\Delta$NNLO$_{\rm GO}$(394), although the difference between the two at the origin is very small. This is consistent with the blue color being prominent in Figure~\ref{fig:16O_HF}. If we now consider the CCD time-averaged density, both interactions lead to the same picture: $\rho_{\rm av, CCD}(r)$ lies above $\rho_{\rm HF}(r)$ at short distances, while they invert at intermediate distances, leading to the red-blue pattern observed in the right panels of Figure~\ref{fig:16O_HF}.

\begin{figure}[htp]
    \centering
    \includegraphics[width=0.49\textwidth]{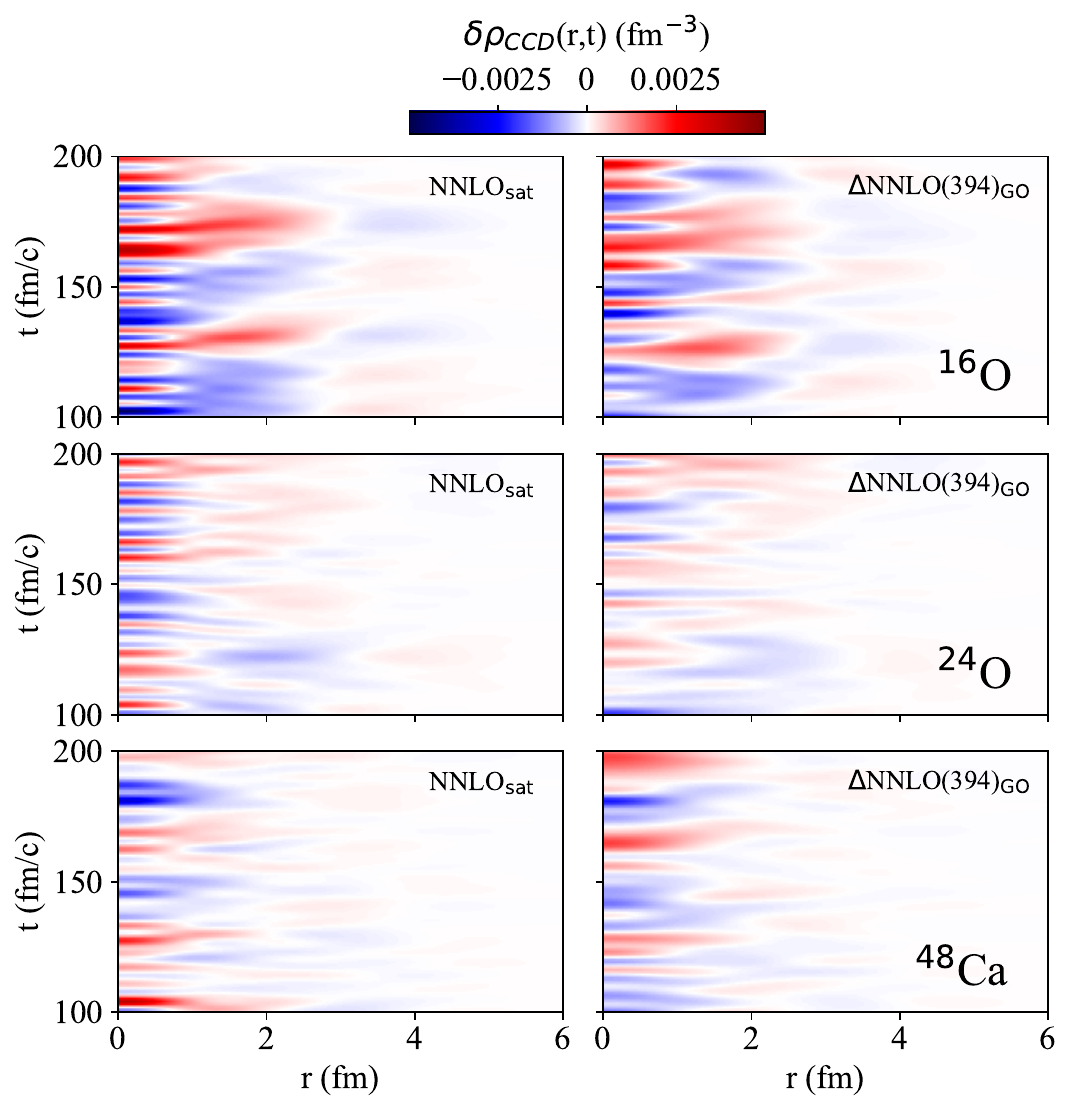}
    \caption{Left panels: Matter density fluctuations of $^{16}$O, $^{24}$O and $^{48}$Ca with respect to the time-averaged density as a function of radial distance $r$ and time $t$ computed with the NNLO$_{\rm sat}$ interaction in the CCD approximation scheme. A model space with $N_{\rm max}=6$ and $\hbar\Omega = 16$ MeV has been employed. Right panels: same for the $\Delta$NNLO$_{\rm GO}$(394) interaction.}
    \label{fig:densities_nmax6}
\end{figure}

\emph{Model space dependence of density fluctuations.--} In this paragraph, we analyse the dependence of density fluctuations on the model space size. To this aim, in Figure~\ref{fig:densities_nmax6}, we show CCD density fluctuations computed with $N_{\rm max} = 6$ for $^{16,24}$O and $^{48}$Ca and both interactions considered in this work.

Also in a smaller model space, we observe the same phenomenon: short-range, short-time fluctuations with an amplitude of around $1\%$ saturation density emerge for all the nuclei under study. Again, shorter periods characterize the results obtained with the harder interaction NNLO$_{\rm sat}$. 

\begin{figure}[htp]
    \centering
    \includegraphics[width=0.49\textwidth]{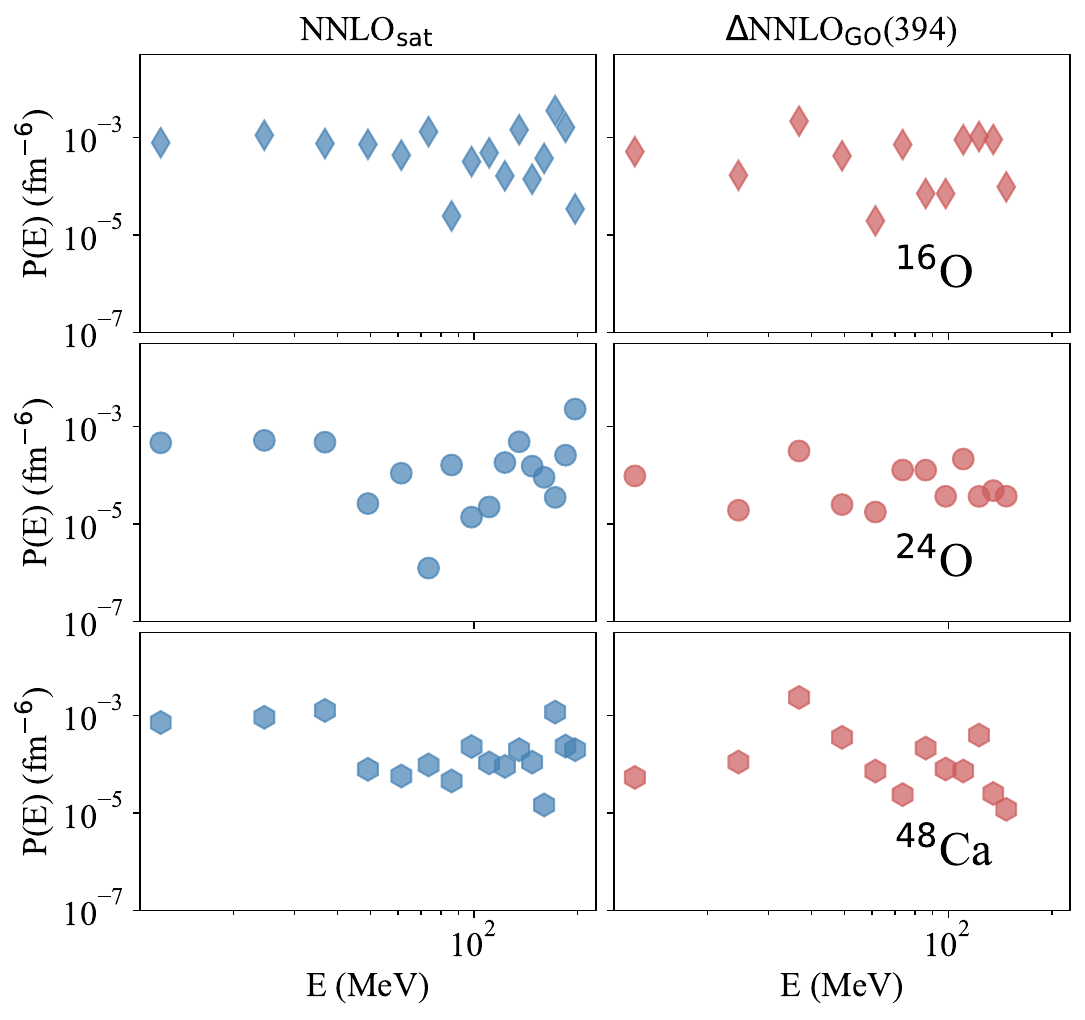}
    \caption{Left panels: Power spectrum of CCD density fluctuations in $^{16}$O (diamonds), $^{24}$O (circles) and $^{48}$Ca (hexagons) computed with the NNLO$_{\rm sat}$ interaction. Right panels: same for the $\Delta$NNLO$_{\rm GO}$(394) interaction. A model space with $N_{\rm max}=6$ and $\hbar\Omega = 16$ MeV has been employed. Both axes employ a logarithmic scale.}
    \label{fig:power_spectra_nmax6}
\end{figure}

In Figure~\ref{fig:power_spectra_nmax6}, we show the corresponding power spectra, obtained with $N_{\rm max} = 6$. Results are qualitatively similar to those obtained with $N_{\rm max} = 8$ in Figure~\ref{fig:power_spectra}: the spectra are almost energy-independent and comparable to white noise. 

\emph{Videos.--} We also produced videos of the density fluctuations from CCD shown in the right panels of Fig.~\ref{fig:48Ca}. The videos start at time $t=0$ and end at time $t=t_{\rm max}=200$~fm$/c$.  
We see that the amplitudes of the fluctuations are particularly large at the beginning (because we start from the Hartree-Fock density) and later become smaller.  For this reason, we choose the starting point for the calculation of the time-averaged density to be $\tau_{\rm min}=100$~fm$/c$.  

\end{document}